\documentclass[epj]{svjour}
\usepackage{graphics}
\begin{document}
\def \nngg {$\nu\bar\nu\gamma(\gamma)$}
\def\GeV{\ifmmode {\mathrm{\ Ge\kern -0.1em V}}\else
                   \textrm{Ge\kern -0.1em V}\fi}%
\def\TeV{\ifmmode {\mathrm{\ Te\kern -0.1em V}}\else
                   \textrm{Te\kern -0.1em V}\fi}%
\def\gravin{\ensuremath{\tilde\mathrm{G}}}%
\def\DELTAM{\ensuremath{\Delta m}}%
\def\Msel{\ensuremath{m_{\tilde\e_\mathrm{L}}}}%
\def\Mse{\ensuremath{m_{\tilde\mathrm{e}}}}%
\def\Mselr{\ensuremath{m_{\tilde\e_\mathrm{L,R}}}}%
\def\Mchi{\ensuremath{m_{\tilde\chi^0_1}}}%
\def\Mchii{\ensuremath{m_{\tilde\chi^0_2}}}%
\def\Mcha{\ensuremath{m_{\tilde\chi^\pm_1}}}%
\def\Msnu{\ensuremath{m_{\tilde\nu}}}%
\def\Msml{\ensuremath{m_{\tilde\mu_\mathrm{L}}}}%
\def\Mstat{\ensuremath{m_{\tilde\tau_2}}}%
\def\MP{\ensuremath{m_{\mathrm{P}}}}%
\def\MG{\ensuremath{m_{\gravin}}}%
\def\susyl#1{\ensuremath{\tilde{#1}_\mathrm{L}}}%
\def\susyr#1{\ensuremath{\tilde{#1}_\mathrm{R}}}%
\def\susylr#1{\ensuremath{\tilde{#1}_\mathrm{L,R}}}%
\def\tanb{\ensuremath{\tan \beta}}%

\title{Searches for New Physics in Photonic Final States at LEP}
\author{Marat Gataullin (on behalf of the LEP Collaborations)
}                     
\offprints{marat@hep.caltech.edu}          
\institute{California Institute of Technology, 
Department of Physics,
Pasadena, CA 91125}
\date{Received: date / Revised version: date}
%
\abstract{
A brief review of  searches for physics beyond the Standard Model in
photonic final states at LEP is given here. These include searches
for supersymmetry, large extra dimensions and contact interactions.
Recent results from all four LEP experiments are presented,
including improved limits on the new scale of gravity for models with large
extra dimensions and the most precise direct measurement of the number
of light neutrino species.  Status and prospects of the LEP combined
searches are also discussed.
} 
\authorrunning{Marat Gataullin}
\maketitle
%
\section{Introduction}
\label{intro}

Photonic final states were 
produced at LEP via the 
reactions $\mathrm{e}^+\mathrm{e}^-\rightarrow\nu
\bar{\nu} \gamma (\gamma) $
 and $\mathrm{e}^+\mathrm{e}^-\rightarrow\gamma  \gamma (\gamma)$,
 leading to two distinct topologies:
single- and multi-photon events with missing energy and events
with collinear photons (di-photons), respectively.
These  experimental signatures are also predicted by a wide variety of
theories with physics  beyond the Standard Model.
Single- and multi-photon events can be used in direct searches
for  new neutral particles, such as graviton production in models
with extra dimensions and neutralino and gravitino production processes
in supersymmetry. 
Whereas in the di-photon topology,  New Physics can manifest itself through
deviations in the measured total and differential cross sections.

 Results reviewed here are based on studies of the photonic final states by 
the four LEP collaborations, ALEPH, DELPHI, L3 and OPAL,
using the highest centre-of-mass energy, $\sqrt{s}$, and 
luminosity LEP data samples  collected in 1998-2000 at
$\sqrt{s} = 189-208 \GeV$ with an integrated luminosity of
about 650~pb$^{-1}$ per experiment.

\section{Single- and Multi-Photon Signatures}
\label{singlephot}
\subsection{Neutrino Production}
\label{neutrino production}

\begin{figure}
\begin{center}
\resizebox{0.4\textwidth}{!}{%
  \includegraphics{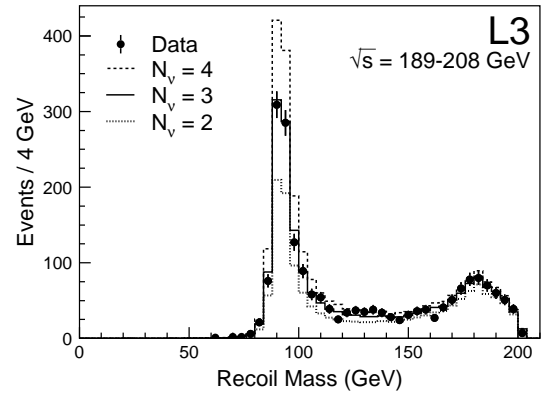}
}
\end{center}
\caption{
The recoil mass spectrum of the
 single- and multi-photon events selected by L3 
compared to the expected spectra for 
$ N_\nu = 2,3$ and 4.
}
\label{fig:1}       
\end{figure}

In the Standard Model of the electroweak interactions the
reaction $\mathrm{e}^+\mathrm{e}^-\rightarrow\nu
\bar{\nu} \gamma (\gamma) $ proceeds through $s$-channel Z exchange and
$t$-channel W exchange, where the photons are radiated mainly
from the incoming electrons and positrons. 
The distribution of the recoil mass to the photon system, $M_{rec}$, is
expected to peak around the Z mass in the $s$-channel, whereas 
photons from the $t$-channel W exchange are expected to have a relatively 
flat energy distribution, peaked at low energies.
 A typical selection (L3) of single- and multi-photon events at LEP requires 
no charged tracks and 
the transverse momentum of the photon system, $P_t^\gamma$,
greater than $0.02\sqrt{s}$.
 The purity of the selected
$\nu\bar{\nu} \gamma (\gamma)$ sample is 99\% and the selection efficiency
is estimated to be about 71\%~\cite{l3sg}.

The expected total and differential
cross sections depend on the number of light neutrino families, $N_\nu$.
The  $M_{rec}$ spectrum of the single- and multi-photon events
selected by the L3 experiment is shown in Figure~\ref{fig:1} together
with the expectations for $N_\nu=2,3$ and 4. 
To determine  $N_\nu$ a  maximum likelihood fit 
is performed
to the two-dimensional
distribution of $M_{rec}$ {\it vs.} $|\cos\theta_\gamma|$. 
Including lower energy data, 
$N_\nu$ is
determined to be~\cite{l3sg}
$$ N_\nu = 2.98 \pm 0.05 (stat) \pm 0.04 (syst).$$
This result is more precise than the
present world average of measurements relying on the single-photon
method~\cite{hagiwara}. ALEPH~\cite{alephsg} and DELPHI~\cite{delphisg}
have performed similar measurements of $N_\nu$, which are also consistent
with the Standard Model value of $N_\nu=3$.

\subsection{Searches for Extra Dimensions}

 Models with large extra dimensions~\cite{qgrav1} predict a
gravity scale, $M_D$, as low as the electroweak scale, naturally
solving the hierarchy problem. Gravitons, G, are then produced in
$\mathrm{e}^+\mathrm{e}^-$
 collisions through the process $\mathrm{e}^+\mathrm{e}^-\rightarrow
 \gamma$G, and escape
detection, leading to a single-photon signature.

\begin{figure}
\begin{center}
\resizebox{0.41\textwidth}{!}{%
  \includegraphics{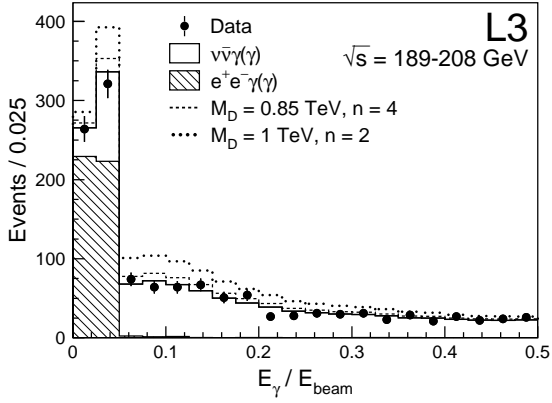}
}
\end{center}
\caption{
 Distribution
of the ratio of the photon energy to the beam energy,
 for the single-photon sample selected by L3. 
Expected signals from the reaction $\mathrm{e}^+\mathrm{e}^-\rightarrow
 \gamma$G  are also shown.}
\label{fig:2}       
\end{figure}

All LEP experiments have performed searches
for this  reaction using
 selected samples 
of single-photon events. 
Since the photon energy spectrum from the graviton production
 is expected to be soft, 
the L3 experiment has also extended its standard
single-photon selection to accept photons with  $P_t^\gamma$ as low
as 0.008$\sqrt{s}$~\cite{l3sg}. 
In the low $P_t^\gamma$ region the Standard Model background 
is increased due to the reaction
$\mathrm{e}^+\mathrm{e}^- \rightarrow \mathrm{e}^+\mathrm{e}^-\gamma(\gamma)$, 
where both electrons have a very low polar angle
and cannot be detected.
 However, this effect is compensated by a significant increase in
 the accepted signal cross section.
Effects of extra dimensions on the L3 photon energy spectrum
are shown in Figure~\ref{fig:2}.

A good agreement with the Standard Model expectation is observed and
limits on the parameter $M_D$ are derived from a fit to the photon energy and
polar angle distributions.
Recent limits obtained at LEP are detailed in Table~1. These  are 
the best current collider limits for the number of extra dimensions below
6.

\begin{table}[htb]
\caption{Lower limits at 95\% C.L. on the gravitational scale, $M_D$,
as a function of the number of extra dimensions, $n$, obtained
by ALEPH~\cite{alephsg}, DELPHI~\cite{delphisg} (preliminary)
 and L3~\cite{l3sg}.}
\label{tab:1}       
\begin{center}
\begin{tabular}{lccc}
\hline\noalign{\smallskip}
 & ALEPH~ & DELPHI & ~~L3~~ \\
$n$ &  \multicolumn{3}{c}{ $M_D$ (\TeV) } \\
\noalign{\smallskip}\hline\noalign{\smallskip}
2   & 1.26 & 1.36 & 1.50 \\
4   & 0.77 & 0.82 & 0.91 \\
6   & 0.57 & 0.59 & 0.65 \\
\noalign{\smallskip}\hline
\end{tabular}
\end{center}
\end{table}

The presence of the brane  in theories with extra 
dimensions creates additional degrees of freedom. 
Brane fluctuations in 
the extra-space directions
 would then manifest themselves as new stable 
particles,
called ``branons'', $\pi_{Br}$~\cite{morat_bran}.
If the brane tension is below the gravity scale,
branons  can be detected at LEP via the
reaction $\mathrm{e}^+\mathrm{e}^- \rightarrow \pi_{Br}\pi_{Br}\gamma$, 
leading to a single-photon signature.
 The signal properties are
similar to those of the graviton production process. 
 L3 has limited branon masses to be above
103~\GeV\ for a scenario with small brane tensions. Alternatively,
under the assumption of light branon masses, brane tensions below 206~\GeV\
are excluded~\cite{l3bran}.

\subsection{Searches for SUSY}

Single- and multi-photon events can be also produced by a variety of processes
predicted in different  models with supersymmetry (SUSY)~\cite{kane}.
These processes  involve production and decays of neutralinos and gravitinos.
No evidence for such models is observed and corresponding limits on SUSY 
parameters  are given in References~\cite{l3sg,alephsg,delphisg}. 
Combined searches are also performed 
by the LEP SUSY Working group~\cite{lepsusy}.

\begin{figure}[htb]
\begin{center}
\resizebox{0.35\textwidth}{!}{%
  \includegraphics{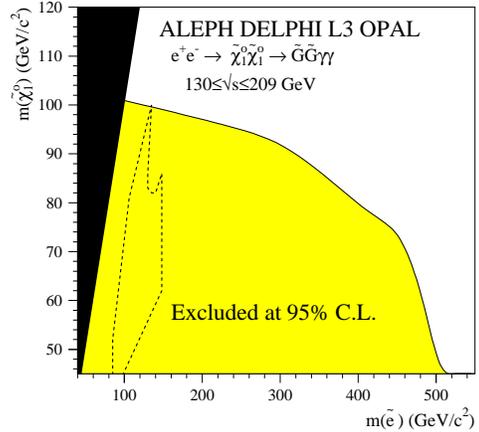}
}
\end{center}
\caption{Region excluded at 95\% C.L. in the \Mchi\ {\it vs.} \Mse\
    plane.  Overlaid is the region favored by the 
interpretation of the CDF event in the scalar
    electron scenario. This result is preliminary.
}
\label{fig:3}       
\end{figure}

   In particular, a search for pair-production of neutralinos, each
decaying into a photon and a gravitino, is motivated by an 
interpretation~\cite{ln96}
of the rare ee$\gamma\gamma$ event observed by CDF~\cite{cdfevent}. 
This reaction, $\mathrm{e}^+\mathrm{e}^- \rightarrow \tilde\chi^0_1 
\tilde\chi^0_1 \rightarrow
\tilde\mathrm{G} \gamma \tilde\mathrm{G} \gamma$, is predicted
 by models with  gauge-mediated SUSY breaking
when the lightest and next-to-lightest supersymmetric particles are
gravitino, $\tilde\mathrm{G}$, and neutralino,  $\tilde\chi^0_1$,
respectively.
The experimental signature of this process is very clean, involving
 events with two energetic acoplanar photons.
No anomalous production of such events has been observed, and Figure~3
shows
the region excluded by LEP  in the 
\Mchi\ {\it vs.} \Mse\  plane, where 
\Mchi\ and \Mse\ are the neutralino and scalar electron masses. 
The above interpretation of the rare CDF event  is now excluded.

\section {Collinear Photons}

Events with  collinear photons are typically selected
by requiring two energetic back-to-back photon candidates and
 no matching charged tracks.
 The cross section of this process has been
measured by all four LEP collaborations~\cite{alephsg,delphi2g,l32g,opal2g}.
The individual measurements have been combined by the LEP Di-photon
Working Group~\cite{lep2g}. 
The measured total cross sections normalized to the QED predictions
are shown in Figure~4. To search for possible signs of new physics
a global fit to the measured total and differential cross sections is
 performed. Good agreement with the Standard Model expectation is observed,
and preliminary LEP combined 
limits have been derived in the context of several New Physics Models,
some of which are described below.

\begin{figure}
\begin{center}
\resizebox{0.48\textwidth}{!}{%
  \includegraphics{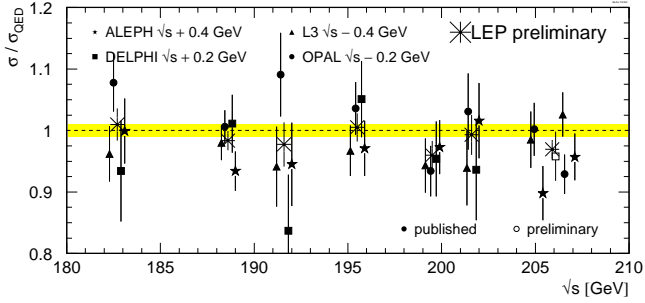}
}
\end{center}
\caption{
The ratio of the measured and QED predicted  cross sections of  the process 
$\mathrm{e}^+\mathrm{e}^-\rightarrow\gamma  \gamma (\gamma)$ 
as a function of $\sqrt{s}$.
}
\label{fig:4}       
\end{figure}

The process $\mathrm{e}^+\mathrm{e}^-\rightarrow\gamma  \gamma (\gamma)$ has
a simple QED description at tree level, and provides a benchmark test of the
QED at  $\mathrm{e}^+\mathrm{e}^-$ colliders. A simple and convenient way 
of parameterizing possible deviations from QED  is the introduction of
the cut-off parameters $\Lambda^{\pm}$. In a similar way, bounds on the
mass scale of  $\mathrm{e}^+\mathrm{e}^-\gamma \gamma$ contact interactions
can be derived in terms of a parameter $\Lambda_7$. The corresponding
 95\% C.L. limits are:
$$  \Lambda^{+} > 392 \GeV,~~ \Lambda^{-} > 364 \GeV,~~
 \Lambda_7 > 837 \GeV.$$

In models with extra dimensions,
photon pair production via virtual graviton exchange 
can interfere with the Standard Model diagrams, leading to 
modifications of the  differential cross section~\footnote{
It should be noted here, that searches for 
manifestations of extra dimensions are performed
  not only in photonic final states but in many other
final state topologies~\cite{mg_ja}.}. 
Deviations from QED can be then
described in terms of a new mass scale, $M_S$, and a  parameter
 $\lambda=\pm 1$, which  gives the sign of the interference.
The derived 95\% C.L. limits are given by:
$$  M_S(\lambda=+1) > 933 \GeV,~ M_S(\lambda=-1) > 1010 \GeV.$$

\section{Conclusions and Discussions}
Searches for New Physics in photonic states have been performed at LEP using the
complete LEP data sample.
These include searches
for supersymmetry, large extra dimensions and deviations from QED.
 No evidence of such models is found.
Constraints on various New Physics theories are 
set by the LEP experiments separately and by preliminary LEP combinations.
Several search strategies and results have been briefly reviewed in this paper. 
Individual references should be consulted for details.

The LEP combinations are expected to be finalized in the near future, and 
an improvement in sensitivity of the LEP combined searches with single-
multi-photon events is expected.
The L3 experiment has also published the selection results
 in the form of tables~\cite{l3sg,l32g}, which can be used to test future
 models involving photonic final states at LEP.

\section{Acknowledgements}
\begin{acknowledgement}
I am very grateful to S.~Mele and H.~Newman for their invaluable 
 help and guidance in preparing this paper.
This work was supported in part by  the 
 U.S. Department of Energy Grant
No.~DE-FG03-92-ER40701.
\end{acknowledgement}

\end{document}